\documentstyle[12pt,a4]{article}

\begin{document}

{\Huge\bf\centerline{P = NP}}
\bigskip
{\Huge\bf\centerline{The Kleene-Rosser Paradox}}

{\Huge\bf\centerline{The Liar's Paradox}}

{\Huge\bf\centerline{\&}}

{\Huge\bf\centerline{A Fuzzy Logic Programming Paradox}}

{\Huge\bf\centerline{$\Longrightarrow$}}

{\Huge\bf\centerline{SAT is (NOT) NP-complete}}

\bigskip\bigskip
{\centerline {Rafee Ebrahim Kamouna}}

{\centerline {Email: rafee102000@yahoo.com}}

{\centerline{submitted to the ACM Transactions on Computation Theory}}

\bigskip
{\centerline {\tt What is a Turing machine?}}

{\centerline {\tt Imeptuous Fire,}}

{\centerline {\tt Syntactico-Semantical!}}

{\centerline {\tt Ice and Desire,}}

{\centerline {\tt {\em Computation} wags on... }}

{\centerline {\em \ \ \ \ \ \ \ \ \ \ \ \ \ \ \ \ \ \ \ \ \ \ \ \ \ \ \ \ \ \ \ \ \ \ \ \ \ \ \ \ \ \ \ \ \ \ \ \ \ \ \ \ \ \ \ \ \ \ \ \ [Turing \`a la ``Romeo \& Juliet"]}}

\bigskip\bigskip
{\bf\centerline{Abstract}}
\bigskip

\noindent After examining the {\bf P} versus {\bf NP} problem against the Kleene-Rosser paradox of the $\lambda$-calculus [94], it was found that it represents a counter-example to NP-completeness. We prove that it contradicts the proof of Cook's theorem. A logical formalization of the liar's paradox leads to the same result. This formalization of the liar's paradox into a computable form is a 2-valued instance of a fuzzy logic programming paradox discovered in the system of [90]. Three proofs that show that {\bf SAT} is (NOT) NP-complete are presented. The counter-example classes to NP-completeness are also counter-examples to Fagin's theorem [36] and the Immermann-Vardi theorem [89,110], the fundamental results of descriptive complexity. All these results show that {\bf ZF$\not$C} is inconsistent.

\break
\noindent{\bf 1. Introduction and the Kleene-Rosser Paradox:}

\bigskip
\noindent This paper examines well-known paradoxes against the fundamental question in complexity theory, i.e. the {\bf P} vs. {\bf NP} problem. The Kleene-Rosser paradox of the inconsistent $\lambda$-calculus discovered in 1935 and a computable formalization of the liar's paradox which is well-known to happen in natural languages. The liar's paradox formalization happens to be a 2-valued special case of a more general multi-valued one. The later being the case of the fuzzy logic programming paradox of the system in [90]. If the {\bf P} versus {\bf NP} problem was ever examined against any of those paradoxes, it would have soon been discovered that it is a straightforward counter-example to NP-completeness.

\bigskip
\noindent Let $L_{\lambda}$ be the language defined by the following function when combined with itself, thus $kk$:
$$k\ =\ (\lambda x.\neg (xx))$$

\noindent one then may deduce
$$kk\ =\ (\lambda x.\neg (xx)) k\ =\ \neg (kk)$$

\noindent Obviously, the language $L_{\lambda}$ is decidable and in {\bf P}. However, it is obvious $L_{\lambda}\not\leq_p$ {\bf SAT}, as how strings which are both ``true" and ``false" can be converted to strings which are either ``true" or ``false". The counter-argument that a Turing machine cannot diagonalize against itself leads to the fact that $L_{\lambda}$ would be a counter-example to the the Church-Turing thesis instead of being a counter-example to NP-completeness. It is implausible to consider such a simply computable language as uncomputable. Also, writing $L_{\lambda}$ as a series of infinite non-halting computations simply ignores that it is programmably implemented and certainly halts. The following proof shows that this paradox results in NP-completeness undefinability when the language $L_{\lambda}$ is assumed to exist.

\bigskip\bigskip
\noindent {\bf Definition 1:} Let $LIAR_{Lang}$ be the class of all languages written in the $LIAR$ logic system and $FLP_{Lang}$ be the class of all languages written in the $FLP$ logic system (defined below), then the class $SySBPD=\{L_{\lambda}:L_{\lambda}\equiv$ The Kleene-Rosser paradox, $L_{\lambda}\in LIAR_{Lang}, L_{\lambda}\in FLP_{Lang}\}$.

\bigskip
\noindent {\bf Definition 2:} Let $M_{\lambda}$ be a program that checks for paradoxes, i.e. a paradox recognizer. A computation $M_{\lambda}$ on $w_{\lambda}\in L_{\lambda}$ prints ``Yes" if $w_{\lambda}$ is an instance of a paradox, i.e. $w_{\lambda}$ = ``True" iff $w_{\lambda}$ = ``False" So:

\begin{enumerate}

\item $M_{\lambda}$ accepts $w_{\lambda}\in L_{\lambda}$ iff $w_{\lambda}$ is paradoxical, otherwise:

\item $M_{\lambda}$ rejects $w_{\lambda}\in L_{\lambda}$ iff $w_{\lambda}$ is satisfiable.

\end{enumerate}

\bigskip
\noindent{\bf Theorem 1.1:} (Main Theorem) {\bf SAT} is NOT NP-complete.

\bigskip
\noindent The line of argumentation of the original proof of {\bf CNF SAT} being NP-complete is as follows as in [21] and quoted from [37]:

\bigskip
\noindent ``Let $A$ be a language in {\bf NP} accepted by a non-deterministic Turing machine $M$: Fix an input $x$. We will create a 3CNF formula $\phi$ that will be satisfiable if and only if there is a proper tableau for $M$ and $x$."

\bigskip
\noindent {\bf Proof:}

\begin{enumerate}

\item Let $M=M_{\lambda}, A=L_{\lambda}, x=w_{\lambda}$.

\item $\Longrightarrow M_{\lambda}$ accepts $w_{\lambda}$.

\item {\bf SAT} is NP-complete.

\item $\Longrightarrow\ [ \forall w_{\lambda}\in L_{\lambda}\ \exists$ a proper tableau for $M_{\lambda}$ and $w_{\lambda}] \Longleftrightarrow [\phi$ is satisfiable].

\item $\Longrightarrow\ \phi$ is satisfiable $\Longleftrightarrow\ M_{\lambda}$ accepts $w_{\lambda}$.

\item But $w_{\lambda}$ is paradoxical, as a paradox.

\item $\Longrightarrow\ \phi$ is satisfiable $\Longleftrightarrow\ w_{\lambda}$ is paradoxical.

\item $\phi$ is satisfiable $\Longleftrightarrow$ ``False".

\item $\phi$ is paradoxical.

\item $\not\exists\ \phi :\phi$ is satisfiable.

\item {\bf SAT} is (NOT) NP-complete.  $\rule{2mm}{2mm}$

\end{enumerate}

\bigskip
\noindent {\bf Theorem 1.2: SAT} is (NOT) NP-complete.

\bigskip
\noindent {\bf Proof:}

\begin{enumerate}

\item {\bf SAT} is NP-complete.

\item $\Longrightarrow\ \forall w_{ij}\in L_i\ \exists\ f(w_{ij})=w_{\bf SAT}\in$ {\bf SAT}.

\item Let $w_{ij}=w_{{\lambda}j}$, then $\exists\ f(w_{{\lambda}}j)=w_{SAT_j}$.

\item $w_{{\lambda}j}$ is ``true" iff ``false" while $w_{SAT_j}$ is either ``true" or ``false".

\item $\not\exists\ f:f(w_{ij})=w_{\bf SAT} \ \forall w_{\lambda}j$.

\item {\bf SAT} is (NOT) NP-complete.   $\rule{2mm}{2mm}$
\end{enumerate}

\bigskip
\noindent {\bf Theorem 1.3: P = NP.}

\bigskip
\noindent {\bf Proof:}

\begin{enumerate}

\item {\bf SAT} is (NOT) NP-complete.

\item $\Longrightarrow$ NP-complete = $\emptyset$.

\item $\Longrightarrow$ {\bf P = NP}.    $\rule{2mm}{2mm}$

\end{enumerate}

\bigskip
\noindent Thus, the Kleene-Rosser paradox known as early as 1935 is sufficient to overturn all NP-completeness results. However, other logical languages may have paradoxical behavior as shown below. Note the misconception of a Turing machine cannot risk contradiction is due to considering it as an encoded integer with no regard to its semantics. Obviously, no integer can form a paradox. A paradox is an {\em absolutely logical} situation which is related to language. Some may transform the Kleene-Rosser paradox as an example of an infinite loop. Concealing this paradox into a physical Turing machine that does not halt would not eliminate it as it is in the language. The liar's paradox which exists in natural language can be easily formalized as below leading to the same above result. Church's $\lambda$-calculus is equivalent to Turing machines among other computational models.

\bigskip\bigskip
{\Huge\bf\centerline{The Syntactico-Semantical Bi-Polar Disorder}}

{\Huge\bf\centerline{Turing Machine Paradox}}

\bigskip
\noindent Since the {\bf P} versus {\bf NP} probnlem has all its roots in the mathematics foundation crisis in the early XX$^{th}$ century, an attempt to examine the reason behind these (negative) results introduce the ``Syntactico-Semantical Bi-Polar Disorder" explained below. The XX$^{th}$ century most important results were re-organized as below:

\begin{enumerate}

\item Self-referential $SySBPD$:

\begin{enumerate}
\item Russell's paradox.

\item The Liar's paradox.

\end{enumerate}

\item G\"{o}del Completeness/Incompleteness $SySBPD$; note the relationship between the proof of his celebrated incompleteness theorem and the Liar's paradox.

\item Turing Decidability/Undecidability $SySBPD$.

\item Finiteness/Infiniteness $SySBPD$: results in finite model theory that succeed infinitely and fail finitely. Most importantly, G\"{o}del's completeness theorem which is:

\begin{enumerate}

\item {\em Positive}: Completeness/Incompleteness $SySBPD$.

\item {\em Negative}: Finiteness/Infiniteness $SySBPD$.

\end{enumerate}

\end{enumerate}

\noindent All these $SySBPD$'s are instances of the {\bf ``Syntactico-Semantical Precedence/Principality Bi-Polar Disorder"}. Note that G\"{o}del completeness theorem is considered a positive results in automated deduction an related areas while considered negative in finite model theory as it fails finitely.

\begin{enumerate}
\item Precedence: syntax definition precedes semantics:

[Syntax $<$ Semantics]$_{Precdence}$.

\item Principality: during computation the input takes various syntactic forms where semantics is principal over syntax in every computation step:

    [Semantics $<$ Syntax]$_{Principality}$.

\item (1) \& (2) $\Longrightarrow$ [Syntax] $<>$ [Semantics], i.e. Bi-Polar Disorder.
\end{enumerate}

\noindent The question:``Are the XX$^{th}$ the only $SySBPD$'s" led to the discovery of all recent results. Now, we have the Syntactico-Semantical Bi-Polar Disorder Turing machine NP-completeness Paradox as:

\bigskip
{\centerline {{\bf SAT} is NP-complete $\Longleftrightarrow$ {\bf SAT} is (NOT) NP-complete}}

\bigskip
\noindent which is simply because:
$$w\ {\tt is\ paradoxical\ \Longleftrightarrow M {\tt accepts}\ w\ \Longleftrightarrow\ A(w)\ {\tt is\ satisfiable}}$$

\bigskip
\noindent where $A(w)$ [21]:
$$A(w)=B\wedge C\wedge D\wedge E\wedge F\wedge G\wedge H\wedge I$$

\bigskip
\noindent and because $P_{s,t}^i$ are propositional variables in $A(w)$
$$P_{1,1}^{i_1}\wedge P_{2,1}^{i_2}\wedge \ldots P_{n,1}^{i_n} {\tt \ is\ satisfiable\ iff}\ w\ {\tt is\ paradoxical}$$

\bigskip
\noindent The reason for the paradox is that Cooks's theorem is still true despite all the results above and below of {\bf SAT} being (NOT) NP-complete. Recent results were obtained solely via logical syntactico-semantical proofs. On the other hand Cook's proof mixes the physical world with the mental world. The formula $A(w)$ in [21] consists of propositional symbols which refer to the physical nature of the Turing machine to prove a property of the set of strings it processes. While the formula is satisfiable from a physical point-of-view, it is not always the case from a logical point-of-view. It is clear that the proof in [21] does not make an account of the meaning of the string $w$ when there is a reference of a computation $M$ on input $w$. $A(w)$ is derived from the machine physical nature during the computation. An example of those physical facts is that if the machine tape head is at the location $k$, then the next computation must be either $k+1$ or $k-1$. This - among many other similar thing - while being a true (physical) property of the machine itself, it may not have implications on the properties of the language being processed. This is the ``Syntactic-Semantical Bi-Polar Disorder Turing machine NP-completeness Paradox" which can be stated more clearly as:

\bigskip

{\centerline {\bf ``A logically satisfiable formula $A(w)$ can always be constructed}}

{\centerline {\bf from the logically paradoxical string $w$"}}

\bigskip
\noindent The source of this contradiction is $w$ has no connection with physics, while $A(w)$ does have. They both meet in the realm of ``Syntax" while they never do in the realm of ``Semantics", hence a syntactico-semantical paradox, which is an irreparable disorder of computation and mathematics. It is possible for a semantic proof to overturn a syntactic one, but not in this case when the proof derives from the physical properties of the non-detrministic Turing machine itself. Obviously, no proof (syntactic or semantic) can overturn any  physical fact, e.g. that if the Turing machine head is at location $k$, then the next computation step must be either at location $k-1$ or at $k+1$. This is a physical fact. The propositional symbols constructed in the proof are mostly derived in this way.

\bigskip
\noindent {\bf 2. The Liar's Paradox:}

\bigskip
\noindent The following theorem proves a formalization of the Liar's paradox in a Prolog style programming language. Thus, self-referential paradoxical languages can be represented in a programming language as well as in the above inconsistent $\lambda$-calculus (recursion vs. self-reference). It is to be noted that self-reference has been removed from first-order logic {\em deliberately a priori} in order to avoid such contradictions. However, its elimination does mean that those contradictions do not exist in the languages (elements) of {\bf P} and {\bf NP}. Consider:

\bigskip
\noindent {\bf P} = $\{L_1,L_2,L_3,\ldots,L_i,L_j,L_k,\ldots\}$.

\bigskip
\noindent Obviously, $L_{\lambda}$ exists as some language in {\bf P} as well as other paradoxical languages like $L\in LIAR$ below. It is of no help to preclude them.

\bigskip
\noindent {\bf Theorem 2.1:} The Liar's Paradox $\equiv$ \{English(John,False)\}.

\bigskip
\noindent {\bf Proof:}

\begin{enumerate}

\item The Liar's Paradox $\equiv$ \{This sentence is False\}.

\item $\Longrightarrow$ \{This sentence is False\} $\equiv$ \{A = A is False\}.

\item $\Longrightarrow$ \{English(John,False)\} $\equiv$ \{A = A is False\}.

\item $\Longrightarrow$ \{This sentence is False\} $\equiv$ \{English(John,False)\}.

\item $\Longrightarrow$ The Liar's Paradox $\equiv$ \{English(John,False)\}.  $\rule{2mm}{2mm}$
\end{enumerate}

\bigskip
\noindent The $LIAR$ logic system has the same $FLP$ [90] syntax and semantics but with truth constants restricted only to two values. Its formulas would look like: $P(f(t),false)$, where $f$ is a recursive function over the recursive term $t$. Simply, the Prolog atom: English(John,False) would be a statement that asserts its own falsehood if and only if it is true, hence a paradox. The first question to address is such statements do exist or do not exist. The liar's paradox do exist in natural language and is well-known for more than two millennia. To assume it is not formalizable in any computable form would never mean that it does not exist. Such an assumption would not stand the test of time against a self-referential question such as {\bf P} vs. {\bf NP} which is itself a question in {\bf NP}. The deliberate elimination of self-reference that may have helped the development of logic would hinder the progress of attacking this question. The reason is that in the development of a logical language, or a class of languages in a logic system, no such a question of whether an infinite class is equal/or not to another infinite class is addressed. Further, in a logical language one is interested to remove any inconsistency a priori. In attacking {\bf P} vs. {\bf NP}, one cannot assume the Kleene-Rosser paradox above does not exist nor ignore its implications. Logical programming languages with paradoxes can be developed like formalizing the Liar's paradox above which happens to be a 2-valued instance of the multi-valued fuzzy logic programming paradox below.

\bigskip\bigskip

\noindent {\bf 3. The Fuzzy Logic Programming Paradox:}

\bigskip
\noindent There is a vast literature with large number of results in both ``Mathematical Fuzzy Logic" and ``Fuzzy Logic Programming", (see the references). Mathematical fuzzy logic systems were developed by Hajek [50-87], Esteva [26,29-35], Godo [37-49] and others. Systems like BL, Lukasiewicz, G\"{o}del and Product logics have been formulated with various rigorous properties and have become standard. Fuzzy logic programming and possibilistic logic programming systems in the works of Godo and Alsinet et al. [1-18], Vojtas et al. [111-115] were developed with large number of soundness and completeness results with interesting properties. Variations as the multi-adjoint logic programming was developed by Medina et al. [98-101]. The huge number of results is clear and of course this is not an exhaustive listing.

\bigskip
\noindent The first use of truth constants in the language syntax first appeared in Pavelka's logic [106] as early as 1979. Before that, truth was expressed only in the language semantics as in Lukasiewicz and Kleene many-valued logics. Pavelka extended Lukasiewicz logic with rational truth constants. Novak [102-105], in his weighted inference systems developed a syntax of pairs: (formula, truth value). Expansions of other logics with truth constants in Esteva et al. 2000, and recently in Esteva et al. 2006 [23-25,27], and Savicky et al. 2006 [108]. In 2007, truth constants appeared in Esteva et al. [28]. The work of Straccia et al. [19,20,96,107,109] in fuzzy description logics employed truth constants as well. So, the idea of having a truth constant in the language syntax is well-established.

\bigskip
\noindent A counter-example to the NP-completeness property written in $FLP$ [90] language is presented. A class of infinite number of languages is characterized - $SySBPD$: the syntactico-semantical bi-polar disorder class including all paradoxical languages of $FLP$ as well as that of the liar's paradox and $\lambda$-calculus. Each element in this class constitutes a counter-example as well. A one-step computation $L$ is introduced to motivate the presentation. Theorem 3.1 establishes the paradox and theorem 4.1 shows that $L$ is {\em decidable} and $L\in P$. Theorem 4.2 establishes the counter-example and shows that {\bf SAT} is NOT NP-complete using the same proof of the Kleene-Rosser paradox.

\bigskip
\noindent First, we recall the fact the syntax of $FLP$ is absolutely classical. All the well-formed formulas of $FLP$ are well-formed formulas of classical logic. However, $FLP$ uses non-classical semantics for the same classical syntax. First, the classical definition of an Herbrand interpretation and an Herbrand model are recalled. Second, it is shown that if truth constants are allowed in the language syntax in the sense of [90], then every Herbrand interpretation of any $FLP$ language is a {\em model} iff it is {\em not a model}, when the case of $FLP$ collapses to classical logic, i.e. $\mu = ``0"$ or $\mu = ``1"$; the $FLP$ paradox is the liar's paradox. This is the ``Syntactico-Semantical Bi-Polar Disorder $FLP$ Paradox". All $LIAR$ well-formed formulas are $FLP$ well-formed formulas. This is why refuting the $FLP$ paradox necessitates refuting its special 2-valued version which is the liar's paradox. It is not easy to refute the liar's paradox nor to show that it is impossible to be formalized in a programming language resulting in the above discussed consequences.

\bigskip
\noindent {\bf Definition 3.1:} Let $L$ be a language over an alphabet $\Sigma$ containing at least one constant symbol. The set $U_L$ of all ground terms constructed from functions and constants in $L$ is called the Herbrand universe of $L$. The set $B_L$ of all ground atomic formulas over $L$ is called the Herbrand base of $L$.

\bigskip
\noindent {\bf Definition 3.2:} The {\em Herbrand interpretation} $I_L$ for a language $L$ is a structure $I_L\equiv <I_c,I_f,I_p>$ whose domain of discourse is $U_L$ where:

\begin{enumerate}

\item $\forall c\in L:c$ is a constant:
$$I_c(c) = c$$.

\item $\forall f\in L:f$ is a function symbol of arity $n$, and $t_1,t_2,\ldots,t_n$ are terms:
$$I_f(f)(t_1,t_2,\ldots,t_n) = f(I(t_1),\ldots,I(t_n))$$

\item $\forall p\in L:p$ is a predicate of arity $n$:
$$I_p(p):B_L\rightarrow\{0,1\}$$

\end{enumerate}

\bigskip
\noindent {\bf Definition 3.3:} The {\em Herbrand interpretation} $I_L$ for a language $L$ is a model iff $I_L:B_L\rightarrow \{1\}\wedge B_L\not\rightarrow \{0\}$.

\bigskip
\noindent Let $L$ be the classical logic program consisting of the single (ground) fact:
$$p(c_1,c_2,\ldots,c_n)\leftarrow$$
\noindent and let $c_n=\mu\in C\subseteq [0,1]$ be a truth constant. If $I_L$ is an Herbrand interpretation for $L$, then $I_L$ is a {\em model} iff it is {\em not a model}. $I_L$ interprets the predicate symbol $p$ (classically) as a relation between the domains from which the n-tuple $(c_1,c_2,\ldots,c_n)$ is extracted. The last member of the tuple $c_n$ is a real number in a countable $C\subseteq [0,1]$. When constant symbols are interpreted in classical semantics, it banishes an argument of a predicate to be the truth constant of the same predicate. $FLP$ non-classical semantics enforces an argument of a predicate to be a truth constant of the same predicate. Semantics of formal languages are enforced in the same way as in natural languages. Since the string ``main" over the Latin alphabet is interpreted differently in English and French (the word ``main" in French means ``hand"). Obviously,

$$Oxford(main) \neq Larousse(main)$$
$$I_{L_{Classical}}\ (p) \not\equiv I_{L_{FLP}}\ (p)$$

\bigskip
\noindent Neither the English people may ask the French to follow Oxford dictionary, nor the French may ask the English to follow Larousse. Forbidding arguments of a predicate to be the truth constant of the same predicate is equally {\bf unacceptable}. Moreover, in the case of the {\bf P} vs. {\bf NP} question, the entire scientific community is pre-occupied with ANY set of strings (a language) that may separate the two classes. Usually, a set of strings in {\bf NP} and not in {\bf P}, hence the question is settled. Let alone the self-referential nature of the question, i.e. {\bf P} vs. {\bf NP} is a question in {\bf NP}. So, if $X$ is the decision problem $X\equiv$ {\bf P} =? {\bf NP}, then $X\in$ {\bf NP}. But classes are (forbidden) to be elements, so such an argument is a meta-mathematical/philosophical one ($X$ is not a valid mathematical object). Just consider an analogy of the question: $x?=y, x\in N, y\in R$. Obviously, this later question is an ill-posed one.

\bigskip
\noindent For the above considerations, the author is not deterred to enforce such semantics on the same syntax of classical logic, then examine the consequences. Forbidding such semantics won't help because both classes contain infinite number of languages. Any method to forbid such semantics can obviously be eliminated with a counter-part to enforce whatever semantics to examine its implications to this long outstanding question. In other words, a counter-argument against $FLP$ non-classical semantics should prove that such languages don't exist at all. The fact that it leads to paradoxical and inconsistent computations never means that these computations are wrong or meaningless. The attached two meta-interpreters work quite well meaningfully from a practical engineering point-of-view. The reason for this is that in a logic programming system, the user is interested in answer substitutions rather than logical consequences as in automatic theorem proving. Cantor's set theory has its famous paradoxes, one can never argue it is wrong, though initially it was controversial. The following theorem proves that languages written in $FLP$ can have interpretations consisting of paradoxical structures.

\bigskip
\noindent{\bf Theorem 3.1:} Let $L$ be the classical logic program consisting of the single (ground) fact:
$$p(c_1,c_2,\ldots,c_n)\leftarrow$$
\noindent and let $c_n=\mu\in C\subseteq [0,1]$ be a truth constant. If $I_L$ is an Herbrand interpretation for $L$, then $I_L$ is a {\em model} iff it is {\em not a model}.

\bigskip
\noindent{\bf Proof:}

\begin{enumerate}

\item $I_L\ \equiv\ <I_c,I_f,I_p>\equiv <I_c,I_p>$.

\item $\Rightarrow I_c(c_1) = c_1$.

\item $\Rightarrow I_c(c_2) = c_2$.

\item $\cdots$

\item $\cdots$

\item $\cdots$

\item $\Rightarrow I_{c_{n-1}}=c_{n-1}$.

\item $\Rightarrow I_c(\mu) = \mu\in [0,1]$.

\item $\Rightarrow I_p\in\{0,1\}$.

\item $\Rightarrow I_L\ \equiv\ <I_c,I_p>$.

\item $\Rightarrow I_L\ \equiv\ <I_c\in [0,1],I_p\in\{0,1\}>$

\item $\Rightarrow$ $\forall I_c\in ]0,1[, I_L$ is a {\em model} iff it is {\em not a model}  $\ \rule{2mm}{2mm}$

\end{enumerate}

\bigskip
\noindent{\bf 4. The $FLP$ Counter-Example to NP-completeness:}

\bigskip
\noindent Consider an $FLP$ program (when $FLP$
is mentioned in this paper, it is meant as defined in [90]). The
definition of a fuzzy atom in $FLP$ is:
$$p(t_1,t_2,\ldots,t_n,\mu)$$

\noindent  Where $\mu\in [0,1]$ is the truth constant. This atom is a classical one despite the weight attached
to it. Consider the $FLP$ program consisting of one fact:

\bigskip
Age-About-21(John,0.9)$\leftarrow$

\bigskip
\noindent The syntax of this program constitutes a well-formed formula of classical logic
programming. Consider the goal:

\bigskip
$\leftarrow$ Age-About-21(John,$\mu$).

\bigskip
\noindent This goal succeeds with two contradictory truth values, namely ``1" and ``0.9".
In computation theory terms, this logic program is a Turing machine $M$
that codes the input string ``$John$" with both ``Yes" and ``No" at the
same time. One for the truth value $\mu = ``1"$ and the other for
$\mu = ``0.9"$, and vice versa. In other words, if the Turing machine
halts in the $q_{accept}$ state, its tape symbols imply that it is
in the $q_{reject}$ state. On the other hand, if it halts in the
$q_{reject}$ state, its tape symbols imply that it is in the
$q_{accept}$ state. This is the $SySBPD$ ``Syntactico-Semantical
Bi-Polar Disorder" paradox. Since the atom in $FLP$ is a classical
one despite the weight attached to it, it is both classical and
fuzzy. So, the $SySBPD$ paradox is due to the fact that:``$p$ is
fuzzy iff $p$ is not fuzzy", or ``$p$ is two-valued iff $p$ is many-valued"
where $p$ is an atom of $FLP$. The
syntax/semantics dichotomy is bi-polarity, and the paradox is the
undesirable disorder.

\bigskip
\noindent{\bf Theorem 4.1:} Let $L$ be the language defined by the above program, then $L$ is {\em decidable} and $L\in P$.

\bigskip

\noindent{\bf Proof:}

\noindent As in [21], $t_M(w)$ denotes the number of steps in the computation of $M$ on input $w$, and $T_M(n)$ the worst case run time of $M$:
$$T_M(n)=max\{t_M(w)|w\in\Sigma^n\}$$

\noindent where $\Sigma^n$ is the set of all strings over $\Sigma$ of length $n$. Let $M$ be the Turing machine associated with the one step computation defined above, clearly:

\begin{enumerate}

\item $T_M(n)=m\in$ N.

\item $\Rightarrow t_M(w)\neq\infty$.

\item $\Rightarrow L$ is decidable.

\item $T_M(n)=m\in$ N $\Rightarrow L\in$ {\bf P}. $\ \rule{2mm}{2mm}$

\end{enumerate}

\noindent The computation $M$ on any input to the above program {\em certainly} {\bf halts} and $L\in$ {\bf P}.

\bigskip
\noindent{\bf Theorem 4.2:} {\bf SAT} is NOT NP-complete.

\noindent Let $L$ over an alphabet $\Sigma$ be the language defined by the $FLP$ program above, then $L$ can NEVER be reduced to {\bf SAT}, hence {\bf SAT} is not NP-complete. The same proof of theorem 1.1 applies.

\bigskip

\noindent{\bf Proof:}

\begin{enumerate}

\item {\bf SAT} is NP-complete.

\item $\Rightarrow L\leq_p$ {\bf SAT}.

\item $\Rightarrow \exists f:\forall x\in L\Leftrightarrow f(x)\in$ {\bf SAT}, [22].

\item $x\in L\Rightarrow \forall x,\ x$ is both {\em accepted} AND {\em rejected} by $M$.

\item $y\in$ {\bf SAT} $\Rightarrow\forall y,\ y$ is either {\em accepted} OR {\em rejected} by $M$.

\item $\Rightarrow\ \not\exists\ f:f(x)=y$.

\item $\Rightarrow L$ can NEVER be reduced to {\bf SAT}.

\item $\Rightarrow$ {\bf SAT} is NOT NP-complete. $\rule{2mm}{2mm}$

\end{enumerate}

\bigskip
\noindent{\bf 5. Why SAT is \underline{NOT} NP-complete}

\bigskip

\noindent Let $L$ be the following Prolog program consisting of the single-fact:
$${\tt Age-About(John,0.9)}\leftarrow$$

\noindent Running this program with any ground goal MUST generate contradictory truth values:

\begin{enumerate}

\item Semantic:``1", or ``0".

\item Syntactic:``0.9".
\end{enumerate}

\noindent There are only two possibilities that have no third:

\begin{enumerate}

\item $L$ has an associated Turing machine $M$, or:

\item $L$ does not have one; the counter-argument that it is impossible for a Turing machine to diagonalize against itself.

\end{enumerate}

\bigskip
\noindent Case (1): $L$ has an associated Turing machine $M$

\begin{enumerate}

\item{\bf SAT Decision Problem:}

\noindent The {\bf SAT} decision function is $R(F,x)$, assigning a truth value $x$ for a Boolean formula $F$.

\begin{enumerate}

\item Input: Boolean Formula.

\item Output: ``1" or ``0".

\end{enumerate}

\item{\bf $L\in LIAR$ and $L\in FLP$ Decision Problems:}

\noindent Both decision problems form a relation that is {\em not} a function $R(F,x,y)$, assigning two distinct truth values, $x$ (AND) $y$ for an $FLP$, or $LIAR$ formula $F$. One truth value is syntactic ``0.9" written above in the program. The other is semantic. Any basic knowledge of logic is sufficient to view both contradictory truth values.

\begin{enumerate}

\item Input: $FLP$ or $LIAR$ Formula.

\item Output:

\begin{enumerate}
\item ``1", $FLP$ semantical truth-value; AND (not or):

\item  ``0.9", syntactical truth-value; or ``0" in the case of the 2-valued $FLP$, i.e.  $L\in LIAR$.

\end{enumerate}

\end{enumerate}

\noindent Those two values are not only irreconcilable, but also irreducible into a (single) truth value.

\end{enumerate}

\noindent Now, the problem is how to write the reduction function $f$ to reduce $L$ to {\bf SAT}:
$$L\leq_p {\tt SAT}$$

\noindent This is a counter-example argument that can be refuted by experiment. If the reader gets angry at this, neither Einstein nor Popper (i.e. ``Testability") would. Who finds himself angry should present the reduction function $f$ reducing $L$ to {\bf SAT} as a refutation to this counter-example. Obviously, this counter-example is just a member of the $SySBPD$ class ($\lambda$-calculus, $FLP$, $LIAR$ and potentially more) of infinite number of languages having the same property.

\bigskip\bigskip
\noindent Case (2): $L$ does not have an associated Turing machine, but $L$ is computable on the von Neumann machine. Then, this is a counter-example to the Church-Turing thesis. The situation becomes:
$$L\in {\bf P}\Longleftrightarrow L\not\in {\bf P}$$

\begin{enumerate}

\item $L\in$ {\bf P}, as it is a one-step computation.

\item $L\not\in {\bf P}$, the class {\bf P} is defined only on Turing machines.

\end{enumerate}

\noindent The claim that $L$ does not have an associated Turing machine should be proved (physically) by building the machine and demonstrating its incapacity compared to the von Neumann machine, i.e. $L$ is not Turing-computable. The skeptic should make a public demonstration of a Turing machine that he claims to be capable of computing everything in history except the above example. In other words, Prolog is (NOT) programmable on Turing machines. Obviously, the above program can be written in all Prolog versions. Thus, he has to prove (experimentally) that PROgramming in LOGic is impossible. A mathematical proof that a Turing machine cannot compute the above program is irrelevant to the physical phenomenon of computation. It would be certainly interesting for everybody to see this machine in public. Of course, not only for the scientific community, but for the whole world.

\bigskip
\noindent Then something must be wrong somewhere. If the Turing machine definition as a tuple in [22] $<\Sigma,\Gamma,Q,\delta>$, then the  counter-argument that it is impossible for a Turing machine to diagonalize against itself definitely assumes that the transition $\delta$ function may not be a logical one. The Turing {\bf SySBPD} machine introduced below emphasizes computable logical functions by assigning logical properties to $\delta$, i.e. $<\Sigma,\Gamma,Q,p(\delta,\mu)>$. It is easy to see that if a Turing machine cannot risk contradiction (as claimed above), then the Turing {\bf SySBPD} may. However, both machines are equivalent with {\bf Turing SySBPD} emphasis of possible logical contradiction.

\bigskip\bigskip
\noindent{\bf 6. Another Proof:}

\bigskip
\noindent This proof is entirely independent of Turing machines. It is easy to see the possibility of such an approach since the {\bf SAT} problem (as well as $FLP$) are logical problems that exist independent of complexity theory. First, the {\bf SAT} and $FLP$ decision problems are defined, then followed by the proof.

\bigskip
\noindent{\bf Definition 6.1:} The {\bf SAT} Decision Problem.

\noindent Let $F$ be a {\bf SAT} formula, the {\bf SAT} computation on $F$ assigns a {\em function} $h$ to $F:h(F)\in\{0,1\}$, thus $h$ is a pair. Either $h=(F,0)$ or $h=(F,1)$. In other words, input string are coded either ``Yes" or ``No".

\bigskip
\noindent{\bf Definition 6.2:} The $FLP$ Decision Problem.

\noindent Let $G$ be an $FLP$ formula, the $FLP$ computation on $G$ assigns a {\em relation} $r$ to $G:r(G)$ is a triple $r(G)=(G,x,y),x\in [0,1],y\in \{0,1\}$ both $x,y$ are non-empty, $x=y$ only when $x,y\in\{0,1\}$, otherwise $x\neq y$; where:

\begin{enumerate}

\item x: syntactic $FLP$ truth value.

\item y: semantic $FLP$ truth value.

\end{enumerate}

\noindent In this case, input strings are coded both ``Yes" and ``No".

\bigskip
\noindent{\bf Theorem 6.1: SAT} is (NOT) NP-complete.

\bigskip
\noindent{\bf Proof:}

\begin{enumerate}

\item {\bf SAT} is NP-complete.

\item $\Longrightarrow\ \forall L$ written in $FLP, L\in$ {\bf P}, {\bf SAT} is NP-complete $\Longrightarrow L\leq_p$ {\bf SAT}.

\item $L\leq_p$ {\bf SAT} $\Longrightarrow r$ is not a relation, but a function, i.e. when the triple must become a pair.

\item $r$ is a relation $\Longrightarrow L\not\leq_p$ {\bf SAT}, contra-positive of 3.

\item $r$ is a relation, by Definition 6.2.

\item $\Longrightarrow$ {\bf SAT} s (NOT) NP-complete. $\ \rule{2mm}{2mm}$

\end{enumerate}

\noindent It is easy to see that infnite-valued $FLP$ is not necessary for the above result and it can be arrived at via only 3-valued $FLP$ as well as 2-valued $FLP$, i.e. the system $LIAR$. The following section presents example programs to demonstrate the invalidity of the counter-argument of the impossibility to write such type of programs.

\bigskip\bigskip
\noindent{\bf 7. The $SySBPD$ Class of Counter-Examples}

\bigskip
\noindent The language $L$ above constitutes a counter-example for the NP-completeness property. In fact, there is
not only one such language but an infinite class of languages, recalling examples in [90] in the context of this paper:

\bigskip

\noindent{\bf Example 7.1 [90]:}

Mature-Student(x,$\mu )\leftarrow$ Student(x),Age-About-21(x,$\mu$)

Age-About-21(John,0.9)$\leftarrow$

Age-About-21(Peter,0.4)$\leftarrow$

Student(John)$\leftarrow$

Student(Peter)$\leftarrow$

\bigskip

\noindent Here, we have three predicate symbols, namely, Student,
Mature-Student and Age-About-21. The n-ary predicate symbol becomes
an n-ary+1 if the predicate is a fuzzy one. This is to allow for the
$\mu$ indicating the membership value. Obviously, Mature-Student and
Age-About-21 are fuzzy predicates. Now, we consider the goal
$\leftarrow$ Mature-Student(John,$\mu$). This will unify the head of
the first rule with unification (x = John, $\mu = \mu$). Thus,
resulting into two subgoals, the first Student(John) which succeeds.
The other subgoal is Age-About-21(John,$\mu$) which succeeds with
the value $\mu = 0.9$ for John. It is obvious that the predicate
Mature-Student leads to the same $SySBPD$ paradox as the
Age-About-21 did above.

\bigskip
\noindent{\bf Example 7.2 [90]:}

Potential-Customer(x,$\mu_1)\leftarrow$ Customer(x),$\mu_1\geq 0.7$

Top-Potential-Customer(x,$\mu_2)\leftarrow$ Customer(x),$\mu_2\geq
0.9$

Good-Credit-Customer(x,$\mu_3)\leftarrow$
Balance-level(x,y,$\mu_3),\mu_3\geq 0.7$

Customer(John) $\leftarrow$

Balance-Level(John,400,0.7)$\leftarrow$

Customer(Richard)$\leftarrow$

Balance-Level(Richard,500,0.8)$\leftarrow$

Consider the goal $\leftarrow$ Good-Credit-Customer(Richard,$\mu$)

\bigskip
\noindent It is obvious that the predicate Good-Credit-Customer
leads to the same $SySBPD$ paradox as the Age-About-21 did above.

\bigskip

\noindent{\bf Example 7.3 [90]:}

R1: $p(x,y,\mu_{p_1})\leftarrow q(x,\mu_{q_1}),r(y,\mu_r)$

R2: $p(x,y,\mu_{p_2})\leftarrow q(x,\mu_{q_2}),s(y,\mu_s)$

R3: $q(m,0.3)\leftarrow$

R4: $r(x,\mu_r)\leftarrow t(x,\mu_t)$

R5: $s(n,1)\leftarrow$

R6: $t(n,0.4)\leftarrow$

\bigskip

\noindent Consider the fuzzy goal $\leftarrow p(m,n,0.3)$ which unifies with
the first fuzzy rule giving the two fuzzy sub-goals, where the
success of each leads to the $SySBPD$ paradox:

\begin{enumerate}

\item $\leftarrow q(m,\mu_{q_1}),\mu_{q_1}\geq 0.3$,

\item $\leftarrow r(n,\mu_r),\mu_r\geq 0.3$.

\end{enumerate}

\bigskip

\noindent The fuzzy subgoal (1) unifies with R3 and succeeds while the second
fuzzy subgoal unifies with R4 and results with another two fuzzy
subgoals with the second being $\mu_r\geq 0.3$ resulting in the goal
$\leftarrow (t,0.3)$ which succeeds when unifying with R6. As a
result, the original goal $\leftarrow p(m,n,0.3)$ succeeds as far as
matching with rule R1 is considered. When matching with rule R2, two
fuzzy subgoals are generated, they are (where the success of each -
again - leads to the $SySBPD$ paradox - and this situation recurs):

\begin{enumerate}

\item $\leftarrow q(m,\mu_{q_2}),\mu_{q_2}\geq 0.3$,

\item $\leftarrow s(n,\mu_s),\mu_s\geq 0.3$.

\end{enumerate}

\noindent The first successfully matches with R3 and the second as well with
R5. So, the original fuzzy goal succeeds in this case.

\noindent Now consider the fuzzy goal $\leftarrow p(m,n,0.2)$ when matching
with R1, two fuzzy subgoals are generated, namely:

\begin{enumerate}

\item $\leftarrow q(m,\mu_{q_1}),\mu_{q_1} \geq 0.2$,

\item $\leftarrow r(n,\mu_r),\mu_r\geq 0.2.$

\end{enumerate}

\noindent The first fuzzy subgoal of (1) $\leftarrow q(m,\mu_{q_1})$ unifies
with R3 giving $\mu_q = 0.3$ and as a result the second fuzzy
subgoal $\mu_q\geq 0.2$ succeeds. For the second fuzzy subgoal
$\leftarrow r(n,\mu_r),\mu_r\geq 0.2$, we have only rule R4 which
unifies successfully resulting in the goal $\leftarrow (t,0.2)$
which succeeds when unifying with R6. As a result, the original
fuzzy goal $\leftarrow p(m,n,0.2)$ succeeds. When matching with R2,
two fuzzy subgoals are generated, namely:

\begin{enumerate}

\item $\leftarrow q(m,\mu_q),\mu_q\geq 0.2$,

\item $\leftarrow s(n,\mu_s),\mu_s\geq 0.2$.

\end{enumerate}

\noindent The first subgoal matches with R3 and succeeds. The second fuzzy
subgoal matches with R5 and succeeds. Now consider a fuzzy goal with
a variable $\mu$, i.e. $\leftarrow p(m,n,\mu)$, matching with R1, we
get:

\begin{enumerate}

\item $\leftarrow q(m,\mu_q),\mu_q\geq \mu$,

\item $\leftarrow r(n,\mu_r),\mu_r\geq \mu$.

\end{enumerate}

\noindent The first matches with R3 and $\mu_q = 0.3$, thus solving $\mu \leq
0.3$. The second will unify with rule R4 then rule R6 returning
$\mu\leq 0.4$. The original goal succeeds with $(\mu\leq
0.3)\wedge(\mu\leq 0.4)$. Thus $\mu\leq 0.3$. When matching with
rule R2, two fuzzy subgoals are generated:

\begin{enumerate}

\item $\leftarrow q(m,\mu_q),\mu_q\geq \mu$,

\item $\leftarrow s(n,\mu_s),\mu_s\geq \mu$.

\end{enumerate}

\noindent The first matches with R3 giving $\mu\leq 0.3$. The second matches
with R5 giving $\mu\leq 1$. The original goal succeeds with
$[(\mu\leq 0.3)\wedge (\mu\leq 1)]\vee [(\mu\leq 0.3)\wedge (\mu\leq
0.4)]$. Thus, $\mu\leq 0.3$. Thus, the $SySBPD$ paradox is generated
and re-generated in this simple program.

\bigskip\bigskip
\noindent{\bf 8. SySBPD Implemented:}

\noindent{\bf 8.1 An $FLP$ Meta-Interpreter: Sun-Unix (IC-Prolog)}

\bigskip
\noindent In this section,
a meta-interpreter is presented to the $SySBPD$ class. The meta-interpreter is implemented in IC-Prolog.
Given the rule:
$$<p_1(x),\mu_{p_1}>\leftarrow <q(x),\mu_{q_1}>.$$

\noindent It can be read declaratively or procedurally:

\begin{enumerate}
\item The declarative reading states that: for a certain value of the
variable $x$, $p_1$ should be true to a level $\mu_{p_1}\geq \mu_{q_1}$.

\item The procedural reading states that: for a fuzzy goal
$\leftarrow <p_1(m),0.3>$ to succeed, the fuzzy subgoal
$\leftarrow <q(m),0.3>$ must succeed. Further, for the fuzzy goal
$\leftarrow <p(m),0.4>$, the fuzzy sub-goal $q(m,0.4)$ must succeed.

\end{enumerate}

\noindent So, as far as execution is concerned, both values of $\mu$ are instantiated
in the fuzzy rule with the same constant level in the goal and then attempt
succeeding the fuzzy sub-goal. Then, using the meta-interpreter,
the rule is rewritten as follows:

$R1:\ <p_1(x),\mu_{p_1}>\leftarrow <q_1(x),\mu_{q_1}>$

as

$R1': p1(X,Mp1):-q(X,Mp1).$

\noindent Now, consider the fuzzy goal $\leftarrow <p_1(m),V>$, where $V$ is a
variable. Now, the system is queried to what maximum level this fuzzy goal can be
satisfied. This is done via the meta-interpreter predicate $solve(A)$ which
becomes $\leftarrow solve(p_1(m,V)).$
The system predicates {\em functor} and {\em arg} are used.

\noindent When rewriting the fuzzy logic programs in IC-Prolog or standard Prolog,
care should be taken as the semantics associated with fuzzy logic programs
are different than that of standard Prolog. For instance, given
the fact $<q(m),0.3>\leftarrow$, in fuzzy logic programming, it is
considered as a fuzzy fact.
$q$ is said to be true to a level $\mu$ where $0<\mu\leq 0.3$. In standard
Prolog, the goal $\leftarrow q(m,0.25)$ would return the answer ``No". So,
to write a fuzzy fact in Prolog, it should be written as:
$$q(m,Mq):- (Mq\leq0.3),(Mq>0)$$

\noindent During execution within the Prolog model, the answers conform to the given
semantics. Now, the extended rules are extended with a factor $f\in [0,1]$ doubting the
rule:
$$<p_1(x),\mu_{p_1}>\leftarrow (0.9) - <q(x),\mu_q>.$$

\noindent For the goal $\leftarrow <p_1(x),0.3>$ to succeed, the fuzzy goal
$\leftarrow q(x,\mu_q)$ must succeed at least with the value
$0.3/0.9$. To do this in standard Prolog,
the fuzzy fact and the fuzzy rule are rewritten as follows:

$$p_1(X,Mp1):- q(X,Mp1).$$
$$q(m,Mq):- (Mq\leq 0.3/0.9),(Mq>0).$$

\noindent which will lead to the intended meaning.

\noindent Now, if the predicate $q$ happens to be in the body of two fuzzy rules with
different $f$ factors, a different rewriting of the facts is required. For
instance,
one obtains the following two rules and two facts:

$R1:\ <p_1(x),\mu_{p_1}>\leftarrow (0.9) - <q(x),\mu_q>$

$R2:\ <p_2(x,y),\mu_{p_2}>\leftarrow (0.7) - <q(x),\mu_q>, <s(Y),\mu_s>.$

$Fact1:\ <q(m),0.3>\leftarrow$

$Fact2: <s(n),0.4>\leftarrow$

\noindent If this fuzzy logic program is rewritten in Prolog, one gets:

$R1':\ p1(X,Mp1):-q(X,Mp1).$

$R2':\ p2(X,Y,Mp2):- q(X,Mp2),s(Y,Mp2).$

and the two fuzzy facts:

$Fact 1':\ q(m,Mq):- (Mq\leq 0.3/0.9),(Mq>0)$

$Fact 2':\ s(n,Ms):- (Ms\leq 0.4/0.7),(Ms>0).$

\noindent If a fuzzy goal matches with $R1'$, then $Fact 1'$, this would be fine. But
if a fuzzy goal matches with $R2'$, the $q$ fuzzy subgoal must have $f=0.7$
not $0.9$. Thus, given the same predicate occurring in the body of two fuzzy rules with
different $f$ factors, it should be renamed when rewriting. As a result,
the predicate $q$ is renamed in $R2$ to $h$, and one obtains two fuzzy
facts $Fact 1'$ and $Fact 2''$ corresponding to Fact 1 in the original
program:

$R1':\ p_1(X,Mp1):- q(X,Mp1).$

$R2':\ p_2(X,Y,Mp2):- h(X,Mp2),s(Y,Mp2).$

$Fact 1':\ q(m,Mq):- (Mq\leq 0.3/0.9),(Mq>0).$

$Fact 1'':\ h(m,Mh):- (Mh\leq 0.3/0.7), (Mh>0).$

$Fact 2':\ s(n,Ms):- (Ms\leq0.4/0.7),(Ms>0).$

\noindent In the following, a code listing for the meta-interpreter is presented
and a rewritten fuzzy logic program in IC-Prolog that was tested with
the results expected from the semantics for fuzzy logic programming.
The $[0,1]$ interval has been assumed as [0,100], i.e. one hundred
increments.

p1(X,Mp1):- q(X,Mp1).

p2(X,Y,Mp2):-q(X,Mp2),s(Y,Mp2).

p3(X,Y,Z,Mp3):- s(Y,Mp3),t(Z,Mp3),q(X,Mp3).

q(m,Mp2):-(Mp2$=<$4/9),(Mp2$>$0).

s(n,Mpr):-(Mpr$=<$3/7),(Mpr$>$0).

t(l,Mpr):-(Mpr$=<$1/2),(Mpr$>$0).

solve(A,0).

solve(A,X) :- X $>$ 0, functor(A,F,N),F=A,arg(N,A,H),var(H),arg(N,A,X),A,!.

solve(A,X) :- X $>$ 0, Z is X - 1, solve(A,Z).

solve(A) :- solve(A,100).

nt(A):-solve(A),functor(A,F,N),arg(N,A,H),Y is 100-H,write(Y).

\noindent The $solve$ predicate finds the threshold if the goal contained variables.
For a negated goal not containing variables the built-in $not$ predicate
would produce the right answer. If the negated goal contained variables, the
$nt$ predicate above gives the threshold.

\bigskip\bigskip
\noindent{\bf 8.2 Meta-Interpreter: PC:Win-Prolog}

\bigskip
\noindent The following are three clauses which form the program in question.
The meta-interpreter will run in conjunction with this program.
This program can be changed and edited each run while the
meta-interpreter is re-usable across different programs.

p1(X,Mp1):- q(X,Mp1).

p2(X,Y,Mp2):-q(X,Mp2),s(Y,Mp2).

p3(X,Y,Z,Mp3):- s(Y,Mp3),t(Z,Mp3),q(X,Mp3).

\noindent The following are clauses to establish the allowable ranges
for truth values in a Prolog syntax.

q(m,Mp2):-(Mp2$=<$4/9),(Mp2$>$0).

s(n,Mpr):-(Mpr$=<$3/7),(Mpr$>$0).

t(l,Mpr):-(Mpr$=<$1/2),(Mpr$>$0).

\noindent Here starts the meta-interpreter: three different predicates:

solve(A) uniary predicate, solve(A,X) binary predicate and nt(A)

\noindent Base predicate to pass the value of zero level without attempting
recursive calls

solve(A,0).

\noindent Base predicate to pass values greater than zero

solve(A,X) :- X $>$ 0, functor(A,F,N),F=A,arg(N,A,H),var(H),arg(N,A,X),A,!.

\noindent Recursive calls to determine the exact levels

solve(A,X) :- X $>$ 0, Z is X - 1, solve(A,Z).

\noindent Initial run of the goal unifies with this clause head

solve(A) :- solve(A,100).

\noindent To produce results for a negated goal:

nt(A):-solve(A),functor(A,F,N),arg(N,A,H),Y is 100-H,write(Y).

\bigskip\bigskip
\noindent{\bf 9. On the avoidability of the Complexity Class:``SySBPD":}

\bigskip
\noindent Overall, the {\em SySBPD} languages would constitute a {\em reduction obstruction}. Obviously, reduction is of central importance in computability and complexity theories. This new complexity class {\em SySBPD} would overlap virtually every complexity class. Its effect is not confined to obstructing reduction only. It would propagate to many results of descriptive complexity. Fagin's theorem [36] as well as the Immermann-Vardi theorem [89,110] are examined after the discovery of this class. However, the second $FLP$ paradox (appearing when considering the cardinality of the valid formulas of the underlying paradoxical system) has far-reaching implications in mathematics outside complexity theory. It turns out that this paradox *proves* the existence of a transfinite cardinal, hence the ``Continuum Hypothesis" \& the ``Axiom of Choice" are false and {\bf ZFC} is inconsistent [92]. Clearly, this inconsistency result affects all of mathematics and mathematical disciplines: physics, computer science, etc, apart from inconsistency results due to NP-completeness and descriptive complexity.

\bigskip
\noindent In fuzzy logic applications, clearly the paradoxical feature of $FLP$ is undesirable. Perhaps that system was never adopted, unless from a practical point-of-view. Theoretically, it is certainly paradoxical. Practically, the meta-interpreter presented above could be used in ``fuzzy expert systems" without any problems. Moreover, fuzzy logic programming can compute even more computable functions. However, its theoretical paradox is avoidable, see [91] and other $FLP$ systems in the references below. Nevertheless, this paradoxical class of languages is (unavoidable) in complexity theory. The reason is that complexity theory is a theory that studies the computational complexity of classes of infinite number of languages. So, even if $FLP$ is ignored, it does not mean that it does not exist. Moreover, it has been demonstrated that the 2-valued $FLP$ paradox is precisely the liar's paradox which is inevitable in natural languages. Perhaps more interestingly, the paradoxical $FLP$ relation occurs in nature. It reconceptualizes the relation between space and time making a quantum theory of gravity possible, the long outstanding question of theoretical physics [93]. As such, a substantial class of paradoxical languages does exist within the robust class {\bf P}. One has the four new computational complexity classes:

\begin{enumerate}

\item $P_{Sys}=\{L:L\in P\cap SySBPD\}$

\item $P_{NonSys}=\{L:L\in P,L\not\in SySBPD\}$

\item $NP_{Sys}=\{L:L\in NP\cap SySBPD\}$

\item $NP_{NonSys}=\{L:L\in NP,L\not\in SySBPD\}$

\end{enumerate}

\noindent Related to the conventional {\bf P} and {\bf NP} as follows:

\begin{enumerate}

\item {\bf P} $= P_{Sys}\cup P_{NonSys}, P_{Sys}\cap P_{NonSys} = \emptyset$

\item {\bf NP} $= NP_{Sys}\cup NP_{NonSys}, NP_{Sys}\cap NP_{NonSys} = \emptyset$

\end{enumerate}

\noindent The NP-completeness property for {\bf SAT} could be revised in the light of the discovery of the new class as the language which is complete to the new computational complexity class NP$_{NonSys}$:

\bigskip
\noindent {\bf Empirical Observation:} SAT is NP$_{NonSys}$-complete.

\bigskip
\noindent $\forall L\in NP_{NonSys}\ L\leq_p$ SAT $\Longrightarrow$ SAT is NP$_{NonSys}$-complete.

\bigskip
\noindent Other complete languages for other classes should be appropriately modified to exclude any language in the $SySBPD$ class, as it cannot be reduced to such a complete language. For instance {\bf HP} is (NOT) c.e.-complete with similar considerations. Answering any of the following questions, answers the {\bf P} =? {\bf NP} question:

\begin{enumerate}

\item P$_{Sys}$ =? NP$_{Sys}$.

\item P$_{NonSys}$ =? NP$_{NonSys}$; the old question.

\end{enumerate}

\noindent{\bf Observation:} {\bf SAT} $\in$ P$_{NonSys}$ $\Longrightarrow$ P$_{NonSys}$ = NP$_{NonSys}$.

\bigskip
\noindent{\bf 10. Descriptive Complexity:}

\bigskip
\noindent Fundamental results of descriptive complexity must be examined against the $SySBPD$ computational paradoxes. Similar arguments as to NP-completeness can be demonstrated as below.

\bigskip
\noindent {\bf 10.1 Fagin's theorem} [36]: {\bf NP = SO$\exists$}.

\bigskip
\noindent{\bf Theorem 10.1:} {\bf NP$\neq$ SO$\exists$}.

\bigskip
\noindent{\bf Proof:} Let $L$ be a one step paradoxical $FLP$ computation as above.

\begin{enumerate}

\item $L\in$ {\bf P}.

\item $L\in$ {\bf NP}.

\item $L\not\in$ SO$\exists$, $L\in$ SO$\exists\ \Longleftrightarrow\not\exists\ p(t_1,t_2,\ldots,t_n,\mu)\in L$.

\item {\bf NP$\neq$ SO$\exists$}\ \ \ $\rule{2mm}{2mm}$

\end{enumerate}

\noindent{\bf Observation:} {\bf NP$_{NonSys}$ = SO$\exists\ - $ FLP}

\bigskip
\noindent By the notation {\bf SO$\exists\ - $ FLP}, it is meant that atoms of the form $p(t_1,t_2,\ldots,t_n,\mu),$

\noindent $\mu\in [0,1]$ are forbidden, i.e. only purely classical atoms. This observation is a restatement of the old result excluding paradoxical $SySBPD$ languages. {\bf NP$_{Sys}$ =? SO$\exists$} remains an open question.

\bigskip
\noindent{\bf Immermann-Vardi theorem} [89,110]: {\bf P = FO+LFP}.

\bigskip
\noindent{\bf Theorem 10.2:} {\bf P $\neq$ FO+LFP}.

\bigskip
\noindent{\bf Proof:} Let $L$ be a one step paradoxical $FLP$ computation as above.

\begin{enumerate}

\item $L\in$ {\bf P}.

\item $L\not\in$ FO+LFP, $L\in$ FO+LFP $\Longleftrightarrow\not\exists\ p(t_1,t_2,\ldots,t_n,\mu)\in L$.

\item {\bf P $\neq$ FO+LFP}\ \ \ $\rule{2mm}{2mm}$

\end{enumerate}

\noindent{\bf Observation:} {\bf P$_{NonSys}$ = [FO - FLP]+LFP}. The question P$_{Sys}$ =? FO+LFP remains open. Similar arguments hold for other descriptive complexity results over the computational complexity hierarchy.

\bigskip
\noindent{\bf Theorem 10.3:} {\bf ZFC} is inconsistent.

\bigskip
\noindent{\bf Proof:}

{\centerline {[Cook's Theorem [21] $\bigwedge$ Theorems 1.1, 4.2, 6.1]}}

{\centerline {$\bigvee$}

{\centerline {[Fagin's Theorem $\bigwedge$ Theorem 10.1]}

{\centerline {$\bigvee$}

{\centerline {[Immermann-Vardi Theorem $\bigwedge$ Theorem 10.2]}

\bigskip
{\centerline {$\Longrightarrow$}

{\centerline {{\bf ZFC} is inconsistent \ \ \ $\rule{2mm}{2mm}$}

\bigskip\bigskip
{\Huge\bf\centerline{SySBPD Implications}}

{\Huge\bf\centerline{The P versus NP Problem}}

\bigskip\bigskip

\noindent The problem certainly survives the $SySBPD$ class of counter-examples to the NP-completeness property. However, a polynomial-time algorithm for {\bf SAT} no longer implies {\bf P} = {\bf NP}. Nor the non-existence of such an algorithm would imply $P\neq NP$. In its basic informal definition:``Whether easy recognition of a solution implies easy finding one", the problems survives as it always had been. However, the precise definition of the class {\bf P} is divided into two (disjoint) classes $P_{SySBPD}$ and $P_{NonSySBPD}$, written simply as $P_{SyS}$ and $P_{NonSyS}$:

\bigskip
\noindent ${\bf P}=\{L|L=L(M)$ for some Turing machine $M$ which runs in polynomial time$\}$

\bigskip
\noindent ${\bf P} = P_{SyS}\cup P_{Non_{SyS}}$

\bigskip\bigskip
\noindent $P_{SyS} = \{L|L=L[M]$ for some Turing machine $M$ which runs in polynomial time$\}$, where $L[M]$ denotes $M$ accepts $L$ iff $M$ rejects it.

\bigskip\bigskip
\noindent $P_{NonSyS} = \{L|L=L(M)$ for some Turing machine $M$ which runs in polynomial time$\}$, where $L(M)$ denotes $M$ accepts $L$ and strictly does not reject it.

\bigskip\bigskip

\noindent The $SySBPD$ could have members across the entire arithmetic hierarchy. Since the fuzzy logic programs [90] are classical, they have the complete Turing hierarchy computational capability. The usual hierarchy nicely presented in [87] MUST be augmented with the class $SySBPD$, resulting in lots of class separation questions. It is obvious that the counter-example to the NP-completeness property is also a counter-example to c.e.-completeness. The same proof above showing {\bf SAT} not to be NP-complete can be used to prove that HP is NOT c.e.-complete. However, it is undecidable. The new complexity hierarchy - incorporating the $SySBPD$ class describes computable languages on the {\bf Turing SySBPD} machine. While the Turing machine had only two halting states: $q_{accept}$ and $q_{reject}$, the {\bf Turing SySBPD} machine is a Turing machine that has the following halting states:

\begin{enumerate}

\item $q_{accept}$: $M$ halts in $q_{accept}$ and only $q_{accept}$, i.e. no paradoxical halting.

\item $q_{reject}$: $M$ halts in $q_{reject}$ and only $q_{reject}$, i.e. no paradoxical halting.

\item $q_{SySBPD}$. $M$ halts in the state $q_{SySBPD}$ when it halts in $q_{accept}$ iff it halts in $q_{reject}$, i.e. $M$ {\em halts paradoxically}.

\end{enumerate}

\noindent The above (reviewed) definition of the class {\bf P} is on the {\bf Turing SySBPD} machine. The question {\bf P =? NP} has the following possibilities:

\begin{enumerate}

\item {\bf P = NP}.

\item {\bf P}$\ \neq\ {\bf NP}$.

\item {\bf P = NP}$\  \wedge\ $ {\bf P} $\neq$ {\bf NP}.

\item Formally Independent.

\item Both Independent and Dependent.

\end{enumerate}

\noindent However, the empirical non-existence of polynomial-time algorithms for the used to be NP-complete problems would still associate the property with intractability. Nevertheless, the existence or non-existence of such algorithms would not resolve the {\bf P} vs. {\bf NP} problem.

\bigskip\bigskip
\noindent{\bf References:}
\begin{enumerate}

\item T. Alsinet, L. Godo:``Adding similarity-based reasoning capabilities to a Horn fragment of possibilistic logic with fuzzy constants". Fuzzy Sets and Systems 144(1): 43-65 (2004)
2003.

\item T. Alsinet, C. Ans\'{o}tegui, R. Béjar, C. Fern\'{a}ndez, F. Manyà:``Automated monitoring of medical protocols: a secure and distributed architecture". Artificial Intelligence in Medicine 27(3): 367-392 (2003).

\item T. Alsinet, R. Béjar, A. Cabiscol, C. Fern\'{a}ndez, F. Manyà:``Minimal and Redundant SAT Encodings for the All-Interval-Series Problem. CCIA 2002: 139-144.

\item T. Alsinet, L. Godo, S. Sandri:``Two formalisms of extended possibilistic logic programming with context-dependent fuzzy unification: a comparative description". Electr. Notes Theor. Comput. Sci. 66(5): (2002).

\item T. Alsinet, L. Godo:``Towards an automated deduction system for first-order possibilistic logic programming with fuzzy constants". Int. J. Intell. Syst. 17(9): 887-924 (2002)
2001.

\item T. Alsinet, L. Godo: ``A Proof Procedure for Possibilistic Logic Programming with Fuzzy Constants". ECSQARU 2001: 760-771
2000.

\item T. Alsinet, R. Béjar, C. Fernandez, F. Manyà:``A Multi-agent system architecture for monitoring medical protocols". Agents 2000: 499-505.

\item T. Alsinet, L. Godo:``A Complete Calcultis for Possibilistic Logic Programming with Fuzzy Propositional Variables". UAI 2000: 1-10
1999.

\item T. Alsinet, L. Godo, S. Sandri:``On the Semantics and Automated Deduction for PLFC, a Logic of Possibilistic Uncertainty and Fuzziness". UAI 1999: 3-12.

\item T. Alsinet, F. Manyà, J. Planes:``Improved Exact Solvers for Weighted Max-SAT". SAT 2005: 371-377
2004.

\item T. Alsinet, F. Manyà, J. Planes:``A Max-SAT Solver with Lazy Data Structures". IBERAMIA 2004: 334-342.

\item T. Alsinet, C. I. Chesnevar, L. Godo, G. R. Simari: ``A logic programming framework for possibilistic argumentation: Formalization and logical properties". Fuzzy Sets and Systems 159(10): 1208-1228 (2008)

\item T. Alsinet, C. I. Chesnevar, L. Godo, G. R. Simari: ``A logic programming framework for possibilistic argumentation: Formalization and logical properties", Fuzzy Sets and Systems 159(10): 1208-1228 (2008).

\item T. Alsinet, L. Godo:``Adding similarity-based reasoning capabilities to a Horn fragment of possibilistic logic with fuzzy constants". Fuzzy Sets and Systems 144(1): 43-65 (2004)

\item T. Alsinet, L. Godo, S. Sandri: Two formalisms of extended possibilistic logic programming with context-dependent fuzzy unification: a comparative description. Electr. Notes Theor. Comput. Sci. 66(5): (2002)
\item T. Alsinet, L. Godo: Towards an automated deduction system for first-order possibilistic logic programming with fuzzy constants. Int. J. Intell. Syst. 17(9): 887-924 (2002)

\item T. Alsinet, L. Godo: A Proof Procedure for Possibilistic Logic Programming with Fuzzy Constants. ECSQARU 2001: 760-771

\item T. Alsinet, L. Godo: A Complete Calcultis for Possibilistic Logic Programming with Fuzzy Propositional Variables. UAI 2000: 1-10

\item F. Bobillo and U. Straccia:``On Qualified Cardinality Restrictions in Fuzzy Description Logics under Lukasiewicz semantics". In Proceedings of the 12th International Conference on Information Processing and Management of Uncertainty in Knowledge-Based Systems, (IPMU-08), 2008.

\item F. Bobillo and U. Straccia:``fuzzyDL:An Expressive Fuzzy Description Logic Reasoner". In Proceedings of the 2008 International Conference on Fuzzy Systems (FUZZ-08).

\item S. Cook: ``The complexity of theorem proving procedures", Proceedings of the Third Annual ACM Symposium on Theory of Computing, 151–158, pdf version available at the blog of Prof Richard Lipton:
http://rjlipton.wordpress.com/
cooks-paper.

\item S. Cook: ``P versus NP, Official Problem Description", www.claymath.org, 2004.

\item F. Esteva, L. Godo:``Putting together Lukasiewicz and product logic", Mathware and Soft Computing 6:219:234, 1999.

\item F. Esteva, L. Godo, P. Hajek and M. Navara:``Residuated Fuzzy Logics with an Involutive Negation", Archive for Math. Log., 39: 103-124.

\item F. Esteva, L. Godo:``Monoidal t-norm Based Logic", Fuzzy Sets and Systems, 124:271-288, 2001.

\item F. Esteva, L. Godo, P. H\'{a}jek, F. Montagna:``Hoops and Fuzzy Logic". J. Log. Comput. 13(4): 532-555 (2003)

\item F. Esteva, L. Godo and C. Noguera:``On Rational Weak Nilpotent Minimum Logics", J. Multiple-Valued Logic and Soft Computing, 2006.

\item F. Esteva, J. Gispert, L. Godo, C. Noguera:``Adding Truth-Constants to Logics of Continuous t-norms: Axiomatization and Completeness Results", Fuzzy Sets and Systems, 158:597-618, 2007.

\item F. Esteva, L. Godo: Towards the Generalization of Mundici's Gamma Functor to IMTL Algebras: The Linearly Ordered Case. Algebraic and Proof-theoretic Aspects of Non-classical Logics 2006: 127-137

\item F. Esteva, L. Godo, F. Montagna: Equational Characterization of the Subvarieties of BL Generated by t-norm Algebras. Studia Logica 76(2): 161-200 (2004)

\item F. Esteva, L. Godo, F. Montagna: Axiomatization of Any Residuated Fuzzy Logic Defined by a Continuous t-norm. IFSA 2003: 172-179

\item F. Esteva, L. Godo, P. H\'{a}jek, F. Montagna: Hoops and Fuzzy Logic. J. Log. Comput. 13(4): 532-555 (2003)

\item F. Esteva, J. Gispert, L. Godo, F. Montagna: On the Standard and Rational Completeness of some Axiomatic Extensions of the Monoidal T-norm Logic. Studia Logica 71(2): 199-226 (2002)

\item  F. Esteva, L. Godo: On Complete Residuated Many-Valued Logics with T-Norm Conjunction. ISMVL 2001: 81-

\item F. Esteva, L. Godo: Monoidal t-norm based logic: towards a logic for left-continuous t-norms. Fuzzy Sets and Systems 124(3): 271-288 (2001)

\item R. Fagin. Generalized First-Order Spectra and Polynomial-Time Recognizable Sets. Complexity of Computation, ed. R. Karp, SIAM-AMS Proceedings 7, pp. 27-41. 1974.

\item L. Fortnow: Computational Complexity blog,

http://oldblog.computationalcomplexity.org/archive/2003-01-19-archive.html.

\item L. Godo, P.etr H\'{a}jek:``Fuzzy inference as deduction". Journal of Applied Non-Classical Logics 9(1), 1999.

\item L. Godo, P. H\'{a}jek, F. Esteva: A Fuzzy Modal Logic for Belief Functions. IJCAI 2001: 723-732

\item L. Godo, P. H\'{a}jek, F. Esteva: A Fuzzy Modal Logic for Belief Functions. Fundam. Inform. 57(2-4): 127-146, 2003.

\item L. Godo, S. Sandri: Special Issue on the Eighth European Conference on Symbolic and Quantitative Approaches to Reasoning with Uncertainty (ECSQARU 2005). Int. J. Approx. Reasoning 45(2): 189-190 (2007)

\item L. Godo, E. Marchioni: Coherent Conditional Probability in a Fuzzy Logic Setting. Logic Journal of the IGPL 14(3): 457-481 (2006)

\item L. Godo: Symbolic and Quantitative Approaches to Reasoning with Uncertainty, 8th European Conference, ECSQARU 2005, Barcelona, Spain, July 6-8, 2005, Proceedings Springer 2005

\item L. Godo, J. Puyol-Gruart, J. Sabater, V. Torra, P. Barrufet, X. Fàbregas: A multi-agent system approach for monitoring the prescription of restricted use antibiotics. Artificial Intelligence in Medicine 27(3): 259-282 (2003)

\item L. Godo, P. H\'{a}jek, F. Esteva: A Fuzzy Modal Logic for Belief Functions. Fundam. Inform. 57(2-4): 127-146 (2003)

\item L. Godo, R. O. Rodriguez: Graded Similarity-Based Semantics for Nonmonotonic Inferences. Ann. Math. Artif. Intell. 34(1-3): 89-105 (2002)

\item L. Godo, P. H\'{a}jek, F. Esteva: A Fuzzy Modal Logic for Belief Functions. IJCAI 2001: 723-732

\item L. Godo, A. Zapico: On the Possibilistic-Based Decision Model: Characterization of Preference Relations Under Partial Inconsistency. Appl. Intell. 14(3): 319-333 (2001)

\item L. Godo, R. L. de M\'{a}ntaras, J. Puyol-Gruart, C. Sierra: Renoir, Pneumon-IA and Terap-IA: three medical applications based on fuzzy logic. Artificial Intelligence in Medicine 21(1-3): 153-162 (2001)

\item L. Godo, V. Torra: Extending Choquet Integrals for Aggregation of Ordinal Values. International Journal of Uncertainty, Fuzziness and Knowledge-Based Systems 9(Supplement): 17-31 (2001)

\item P. H\'{a}jek:``Metamathematics of Fuzzy Logic", Trends in Logic, Kluwer Academic Publishers, Dordrecht, Vol. 4, 308 pp., 1998.

\item P. H\'{a}jek: On arithmetical complexity of fragments of prominent fuzzy predicate logics. Soft Comput. 12(4): 335-340 (2008)

\item P. H\'{a}jek: Complexity of fuzzy probability logics II. Fuzzy Sets and Systems 158(23): 2605-2611 (2007)

\item P. H\'{a}jek: On witnessed models in fuzzy logic. Math. Log. Q. 53(1): 66-77 (2007)

\item P. H\'{a}jek: On witnessed models in fuzzy logic II. Math. Log. Q. 53(6): 610-615 (2007)
2006

\item P. H\'{a}jek: On Fuzzy Theories with Crisp Sentences. Algebraic and Proof-theoretic Aspects of Non-classical Logics 2006: 194-200

\item P. H\'{a}jek: What is mathematical fuzzy logic. Fuzzy Sets and Systems 157(5): 597-603 (2006)

\item P. H\'{a}jek: Computational complexity of t-norm based propositional fuzzy logics with rational truth constants. Fuzzy Sets and Systems 157(5): 677-682 (2006)

\item P. H\'{a}jek: Mathematical Fuzzy Logic - What It Can Learn from Mostowski and Rasiowa. Studia Logica 84(1): 51-62 (2006).

\item P. H\'{a}jek: Logics for Data Mining. The Data Mining and Knowledge Discovery Handbook 2005: 589-602

\item P. H\'{a}jek: On arithmetic in the Cantor-Lukasiewicz fuzzy set theory. Arch. Math. Log. 44(6): 763-782 (2005)

\item P. H\'{a}jek: Making fuzzy description logic more general. Fuzzy Sets and Systems 154(1): 1-15 (2005)

\item P. H\'{a}jek: A non-arithmetical Godel logic. Logic Journal of the IGPL 13(4): 435-441 (2005)

\item P. H\'{a}jek: Arithmetical complexity of fuzzy predicate logics - a survey. Soft Comput. 9(12): 935-941 (2005)

\item P. H\'{a}jek, Jan Rauch, David Coufal, Thomas Feglar: The GUHA Method, Data Preprocessing and Mining. Database Support for Data Mining Applications 2004: 135-153

\item P. H\'{a}jek: A True Unprovable Formula of Fuzzy Predicate Logic. Logic versus Approximation 2004: 1-5

\item P. H\'{a}jek: On generalized quantifiers, finite sets and data mining. IIS 2003: 489-496

\item P. H\'{a}jek: Relations and GUHA-Style Data Mining II. RelMiCS 2003: 163-170

\item P. H\'{a}jek, Martin Holena, Jan Rauch: The GUHA Method and Foundations of (Relational) Data Mining. Theory and Applications of Relational Structures as Knowledge Instruments 2003: 17-37

\item P. H\'{a}jek: Fuzzy Logics with Noncommutative Conjuctions. J. Log. Comput. 13(4): 469-479 (2003)

\item P. H\'{a}jek: Basic fuzzy logic and BL-algebras II. Soft Comput. 7(3): 179-183 (2003)

\item P. H\'{a}jek: Observations on non-commutative fuzzy logic. Soft Comput. 8(1): 38-43 (2003)

\item P. H\'{a}jek, Martin Holena: Formal logics of discovery and hypothesis formation by machine. Theor. Comput. Sci. 292(2): 345-357 (2003)
2002

\item P. H\'{a}jek: Observations on the monoidal t-norm logic. Fuzzy Sets and Systems 132(1): 107-112 (2002)

\item P. H\'{a}jek: A New Small Emendation of Godel's Ontological Proof. Studia Logica 71(2): 149-164 (2002)

\item P. H\'{a}jek: Monadic Fuzzy Predicate Logics. Studia Logica 71(2): 165-175 (2002)
2001

\item P. H\'{a}jek, Z. Hanikova: A Set Theory within Fuzzy Logic. ISMVL 2001: 319-323

\item P. H\'{a}jek: Relations in GUHA Style Data Mining. RelMiCS 2001: 81-87

\item P. H\'{a}jek, John C. Shepherdson: A note on the notion of truth in fuzzy logic. Ann. Pure Appl. Logic 109(1-2): 65-69 (2001)

\item P. H\'{a}jek, Sauro Tulipani: Complexity of Fuzzy Probability Logics. Fundam. Inform. 45(3): 207-213 (2001)

\item P. H\'{a}jek, L. Godo, S. Gottwald: Editorial. Fuzzy Sets and Systems 124(3): 269-270 (2001)

\item P. H\'{a}jek: On very true. Fuzzy Sets and Systems 124(3): 329-333 (2001)

\item P. H\'{a}jek: Fuzzy Logic and Arithmetical Hierarchy III. Studia Logica 68(1): 129-142 (2001)
2000

\item P. H\'{a}jek, Dagmar Harmancov?: A Hedge for G\"{o}del Fuzzy Logic. International Journal of Uncertainty, Fuzziness and Knowledge-Based Systems 8(4): 495-498 (2000)

\item P. H\'{a}jek, Jeff B. Paris, John C. Shepherdson: The Liar Paradox and Fuzzy Logic. J. Symb. Log. 65(1): 339-346 (2000)

\item P. H\'{a}jek, Jeff B. Paris, John C. Shepherdson: Rational Pavelka Predicate Logic Is A Conservative Extension of Lukasiewicz Predicate Logic. J. Symb. Log. 65(2): 669-682 (2000)

\item P. H\'{a}jek, Jan Rauch: Logics and Statistics for Association Rules and Beyond Abstract of Tutorial. PKDD 1999: 586-587

\item P. H\'{a}jek: Ten Questions and One Problem on Fuzzy Logic. Ann. Pure Appl. Logic 96(1-3): 157-165 (1999)

\item N. Immermann, ``Guest Column: Progress in Descriptive Complexity", SIGACT News Complexity Theory Column 49, ACM SIGACT News, September 2003 Vol. 34, No. 3.

\item N. Immerman, Relational queries computable in polynomial time, Information and Control 68 (1–3) (1986) 86–104.

\item R. E. Kamouna: ``Fuzzy Logic Programming", Fuzzy Sets and Systems, 1998.

\item R. E. Kamouna:``Fuzzy Logic Programming Based on $\alpha$-Cuts", Ph.D. thesis, De Montfort University, England, 2003.

\item R. E. Kamouna:``Two Fuzzy Logic Programming Paradoxes Imply Continuum Hypothesis="False" \& Axiom of Choice="False" Imply ZFC is Inconsistent, http://arxiv.org/PS-cache/arxiv/pdf/0807/0807.2543v4.pdf.

\item R. E. Kamouna: ``A Spatio-Temporal Bi-Polar Disorder Quantum Theory of Gravity, A Fuzzy Logic Programming Reconciliation, http://arxiv.org/PS-cache/arxiv/pdf/0806/0806.2947v7.pdf

\item S.C. Kleene and J. B. Rosser, ``The inconsistency of certain formal logics." Ann. of Math., 36:630-636, 1935.

\item S. Krajci, R. Lencses, P. Vojt\'{a}s:``A comparison  of fuzzy and annotated logic programming". Fuzzy Sets and Systems, 144 (2004) 173–192

\item T. Lukasiewicz and U. Straccia:``Managing Uncertainty and Vagueness", in Description Logics for the Semantic Web In Journal of Web Semantics.

\item S. Krajci, R. Lencses, P. Vojt\'{a}s:``A comparison  of fuzzy and annotated logic programming". Fuzzy Sets and Systems, 144 (2004) 173–192

\item J. Medina, M. Ojeda, P. Vojt\'{a}s:``Multi-adjoint logic programming with continuous semantics". In Proc. LPNMR'01. Th. Eiter et al eds. Lecture Notes in Artificial Intelligence 2173, Springer Verlag 2001, 351-364

\item J. Medina, M. Ojeda, P. Vojt\'{a}s:``A procedural semantics for multi-adjoint logic programming. In Proc. EPIA'01, P. Brazdil and A. Jorge eds. Lecture Notes in Artificial Intelligence 2258, Springer Verlag 2001, 290-297

\item J. Medina, M. Ojeda, P. Voj\'{a}s:``A completeness theorem for multi-adjoint logic programming". In Proc. 10th IEEE  Internat. Conf. Fuzzy Systems, IEEE 2001, 1031-1034,

\item J. Medina, M. Ojeda-Aciego, A. Valverde, P. Vojt\'{a}s:``Towards Biresiduated Multi-adjoint Logic Programming". R. Conejo et al Eds. Revised Selected Papers of CAEPIA 2003. Lecture Notes in Computer Science 3040 Springer 2004, 608-617,

\item V. Novak, I. Perfilieva and J. Mockor: ``Mathematical principles of fuzzy logic", Kluwer, Boston/Dordrecht, 1999.

\item V. Novak:``Weighted inference systems", in J. C. Bezdek, D. Dubois and H. Prade (eds.): Fuzzy Sets in Approximate Reasoning and Information Systems. Handbooks of Fuzzy Sets Series, Vol. 3. Kluwer, Boston, 191-241, 1999.

\item V. Novak and I. Perfilieva (eds.):``Discovering the World with Fuzzy Logic; Studies in fuzziness and soft computing", Heidelberg, New York: Physica-Verlag, Vol. 57, 302-304, 2000.

\item V. Nov\'{a}k, S. Gottwald, P. H\'{a}jek: Selected papers from the International Conference "The Logic of Soft Computing IV" and Fourth workshop of the ERCIM working group on soft computing. Fuzzy Sets and Systems 158(6): 595-596 (2007)

\item J. Pavelka:``On Fuzzy Logic I-III. Zeit", Math Logik Grund. Math. 25, 45-52, 119-134, 447-464, 1979.

\item A. Ragone, U. Straccia, T. Di Noia, E. Di Sciascio and F. M. Donini:``Fuzzy Description Logics for Bilateral Matchmaking in e-Marketplaces". In Proceedings of the 16th Italian Symposium on Advanced Database Systems (SEBD-08), 2008.

\item P. Savicky, R. Cignoli, F. Esteva, L. Godo, C. Noguera:``On Product Logic with Truth-constants, Journal of Logic and Computation, Volume 16, Number 2, pp. 205-225(21), Oxford University, 2006.

\item U. Straccia:``Fuzzy Description Logic Programs", in Uncertainty and Intelligent Information Systems, B. Bouchon-Meunier, R.R. Yager, C. Marsala, and M. Rifqi eds. , 2008.

\item M. Y. Vardi, The complexity of relational query languages, in: Proc. 14th ACM Symp. on Theory of Computing, 1982, pp. 137–146.

\item P. Vojt\'{a}s:``Fuzzy logic programming". Fuzzy Sets and Systems. 124,3 (2001) 361-370

\item P. Vojt\'{a}s, T. Alsinet, Ll. Godo:``Different models of fuzzy logic programming with fuzzy unification (towards a revision of fuzzy databases)". In Proc. IFSA'01 Vancouver, IEEE, 2001, 1541-1546,

\item P. Vojt\'{a}s:``Tunable fuzzy logic programming for abduction under uncertainty". In Proc. Workshop Many Valued Logic for Computer Science Applications. European Conference on Artificial Intelligence 98, University of Brighton, 1998, 7 pages

\item P. Vojt\'{a}s. L. Paulak:``Soundness and completeness of non-classical extended SLD-resolution", in Proc. ELP'96 Extended logic programming, Leipzig, ed. R. Dyckhoff et al., Lecture Notes in Comp. Sci. 1050 Springer Verlag, 1996, 289-301.

\item P. Vojt\'{a}s, M. Vomlelov\'{a}:``Transformation of deductive and inductive tasks between models of logic programming with imperfect information", In Proc. IPMU 2004, B. Bouchon-Meunier et al. eds. Editrice Universita La Sapienza, Roma, 2004, 839-846

\end{enumerate}

\break

{\Huge\bf\centerline{A Spatio-Temporal Bi-Polar Disorder }}

{\Huge\bf\centerline{Quantum Theory of Gravity}}

{\Huge\bf\centerline{A Fuzzy Logic Programming Reconciliation}}

{\Huge\bf\centerline{$SySBPD\Longleftrightarrow SpTBPD$}}

{\centerline {Rafee Ebrahim Kamouna}}

\bigskip
{\centerline {\tt What is Gravity?}}

{\centerline {\tt Imeptuous Fire,}}

{\centerline {\tt Space-Temporal!}}

{\centerline {\tt Ice and Desire,}}

{\centerline {\tt {\em The Universe} wags on... }}

{\centerline {\em \ \ \ \ \ \ \ \ \ \ \ \ \ \ \ \ \ \ \ \ \ \ \ \ \ \ \ \ \ \ \ \ \ \ \ \ \ \ \ \ \ \ \ \ \ \ \ \ \ \ \ \ \ \ \ \ \ \ \ \ [Einstein \`a la ``Romeo \& Juliet"]}

\bigskip

{\centerline {\tt What is a Turing machine?}}

{\centerline {\tt Imeptuous Fire,}}

{\centerline {\tt Syntactico-Semantical!}}

{\centerline {\tt Ice and Desire,}}

{\centerline {\tt {\em Fuzzy Logic Programming} goes on... }}

{\centerline {\em \ \ \ \ \ \ \ \ \ \ \ \ \ \ \ \ \ \ \ \ \ \ \ \ \ \ \ \ \ \ \ \ \ \ \ \ \ \ \ \ \ \ \ \ \ \ \ \ \ \ \ \ \ \ \ \ \ \ \ \ [Einstein meets Turing]}

\bigskip
{\bf\centerline{Abstract}}

\noindent A theory of quantum gravity founded on fuzzy logic programming $FLP$ [1] is presented. The connection between space and time of general relativity is re-examined from a logical point-of-view. A one-to-one correspondence between the space/time dichotomy and syntax/semantics of logic was discovered. The Syntactico-Semantical Bi-Polar Disorder nature of $FLP$ ($SySBPD$) naturally expresses the space/time relationship as well as unifying it with quantum mechanics particle/anti-particle dichotomy. The Spatio-Temporal Bi-Polar Disorder ($SpTBPD$) theory makes new predictions that can be tested by experiment, formulates new hypotheses as well as shedding light on previously unexplained observed phenomena, e.g. ``CP violation" and the 720 degrees instead of 360 for an electron to return to its state.

\bigskip\bigskip

\noindent{\bf Introduction:}

\bigskip
\noindent Einstein's general relativity is the most accepted theory of gravity confirmed by experiments and observations. It is mathematically expressed as tensor equations whose solution is Lorentzian manifolds of curved spacetime (Riemannian/Pseudo-Riemannian space). Dirac's equation is the experimentally-verified relativistic quantum mechanics theory that successfully unified quantum mechanics and special relativity (flat spacetime - Minkowski space) whose solution is a wave function. Establishing a theory of quantum gravity remains (undoubtedly) as the theoretical physics outstanding problem for decades [3]. The standard model of particle physics unified all nature fundamental forces except gravity. Having a unified theory of all fundamental forces of nature is obviously a goal longtime sought after. Einstein had famously spent quite a long time in search for a ``Unified Field Theory". This paper presents a {\em fuzzy logic programming (FLP)} reconciliation of the theory of general relativity (Einstein's field equations) and relativistic quantum mechanics (Dirac's equation). This problem can be formulated as: ``If general relativity regards gravity as spacetime and quantum mechanics provides a wave function ($\Psi$ - Dirac's equation) evolution in time, it seems impossible for those two theories to be (mathematically) unified. Attempts include unsuccessful perturbative quantum gravity, string theories culminating in M-Theory but with neither experimental results nor observations [3].

\bigskip
\noindent It was found that the language of ``Fuzzy Logic Programming $FLP$" [1] can {\em naturally} do the job. This is done via re-examination of the logical relationship between space and time. The discovery of a one-to-one correspondence between the space/time dichotomy and that of syntax/semantics made $FLP$ a naturally appealing candidate for this intractable reconciliation; $SySBPD$ vs. $SpTBPD$.

\bigskip\bigskip

\noindent{\bf Philosophical Foundation:}

\bigskip

\noindent $E=mc^2$ implies that $E$ and $m$ are different
(manifestations) ``essences" of the same ``existence". The
``Principality" of the existence over the essence should be obvious.
There can be no ``Principality" for $E$ over $m$ nor vice versa.
This can be called ``Energy/Mass" Principality Bi-Polar Disorder; to
render the term {\em connotating} and its meaning {connotated to}!

\bigskip
\noindent More importantly, the lessons of general relativity dictate that if the Earth gets into the event horizon of a black hole, space and time would swap positions. This implies that space and time are
different manifestations ``essences" of the same ``existence". It
should be self-evident that there is no ``Principality" outweigher
for neither space nor time over one another. This is the
``Spacetime/Timespace Principality Bi-Polar Disorder". $SpTBPD$ exploits $R_{E,m}(E,m)$ vs. $R_{SpT}(space,time)$, where $R_{E,m}(E,m)$ means energy and mass can swap positions in special relativity and $R_{SpT}(space,time)$ means space and time swap positions in general relativity.

\bigskip
\noindent It is easy to see that space and
time are always swapping positions but only completely within a
black hole. Solutions to Einstein's field equations are spacetimes which are Lorentzian manifolds. The tangent vector at any point in the manifold is classified as spacelike (swapping positions spacewise) or timelike (swapping positions timewise) according to the negative/positive value of the manifold's metric [4]. If $(M,g)$ is a Lorentzian manifold (so $g$ is the metric on the manifold $M$) then the tangent vectors at each point in the manifold can be classed into three different types. A tangent vector $X$ is:

\begin{enumerate}

\item {\bf timelike} if $g(X,X)>0$

\item {\bf null} if $g(X,X)=0$

\item {\bf spacelike} if $g(X,X)<0$.

\end{enumerate}

\noindent General relativity [5] is understood as spacetime tells matter how
to move, then in $SpTBPD$ so should timespace. And if matter tells spacetime how to
curve in general relativity, then in $SpTBPD$ it should tell timespace too. The difference between spacetime and timespace:

\begin{enumerate}

\item For a single observer at one point, they are identical.

\item For two observers $A,B$ at two different locations $X,Y$, we have:

Spacetime(A,X) = Timespace(B,Y)

or

Timespace(A,X) = Spacetime(B,Y)

\end{enumerate}

\noindent That is to say, they are reciprocal. This is a corollary of the
Space/Time Principality Bi-Polar Disorder. So, $SpTBPD$ Quantum Theory of Gravity regards gravity as reciprocal spacetime/timespace and quantum mechanics as reciprocal wave functions $\Psi$-(particle)/$\Psi_{BPD}$-anti-particle. $SpTBPD$ regards spacetime geometry as given by the Einstein Field Equations is a result of a
fermion spin-like angular motion (in a Hilbert space) of flat
spacetime and flat timespace. This would justify the dynamic
spacetime geometry. This philosophical interpretation of spacetime
geometry could extend (potentially reconciling) von Neumann
mathematical foundations of quantum mechanics to general relativity.
This is a new hypothesis. Another one is quantum interpretation of the
Big Bang as well as the expansion of the universe. This is due to
the particle/anti-particle view of spacetime/timespace. Spacetime (in-order) is identical to
timespace (disorder). The difference is a matter of state. The Pauli
exculsion principle could be extended from quantum mechanics to gravity in $SpTBPD$.

\bigskip\bigskip
\noindent{\bf Mathematical Formulation: $SySBPD$ vs. $SpTBPD$}

\bigskip

\noindent The following equations relate two solutions of Einstein's equations with another two of Dirac's. Einstein's solutions are two $BPD$-conformally related Lorentzian manifolds (as defined below). Both are related by a $SySBPD$ $FLP$  predicate $Gravity$. Dirac's two solutions are two wave functions, for particles and anti-particles. Both are related by another $SySBPD$ $FLP$  predicate $Quantum$. The two manifolds with both wave functions are related by $\kappa$, the Universe Bi-Polar Disorder Constant, the prediction of $SpTBPD$.

\bigskip

\begin{enumerate}

\bigskip
\item $Equal(\mu_{\Psi},|\Psi| - |\Psi_{BPD}|)\leftarrow Quantum(\Psi,\Psi_{BPD},\mu_{\Psi})$

\bigskip

\item $Equal(\mu_{gravity},\kappa . \mu_{\Psi})\leftarrow Gravity(Lorentz,Lorentz_{BPD},\mu_{gravity})$.

\end{enumerate}

\bigskip
\noindent Where $Gravity$ and $Quantum$ are tertiary $FLP$ predicates as in [1], and $Equal$ is a binary non-fuzzy predicate whose meaning is obvious. $Lorentz$ and $Lorentz_{BPD}$ are any two $BPD$-conformally related Lorentzian solutions of Einstein's equations. Let $g$ be the $Lorentzian$ manifold metric and ${\hat{g}}$ $Lorentz_{BPD}$ manifold metric, they are conformally related if ${\hat{g}}=\Omega^2 g$ (standard definition [3]) and $BPD$-conformally related if ${\hat{g}}= - \Omega^2 g$ (new definition). So, $\mu_{gravity}= \Omega^2$.

\bigskip
\noindent From [3], and for the paper to be self-contained: ``Two metrics $g$ and $\hat{g}$ are conformally related if $\hat{g}=\Omega^2 g$ for some real function $\Omega$ called the {\bf conformal factor}. Looking at the definitions of which tangent vectors are timelike, null and spacelike we see they remain unchanged if we use $g$ or $\hat{g}$. As an example suppose $X$ is a timelike tangent vector with respect to the $g$ metric. This means that $g(X,X)>0$. We then have that $\hat{g}(X,X)=\Omega^2 g(X,X)>0$, so $X$ is a timelike tangent vector with respect to the $\hat{g}$ too. It follows from this that the causal structure of a Lorentzian manifold is unaffected by a conformal transformation."

\bigskip
\noindent A solution to Dirac's equation is the wave function $\Psi$ associated with the quantum system particles and $\Psi_{BPD}$ is the wave function associated with the corresponding anti-particles. It is to be noted that the above $FLP$ rules/equations are not equivalent to their algebraic counterpart:

\bigskip
$\Omega^2=\kappa\ . \mu_{\Psi}=\kappa\ . |\Psi - \Psi_{BPD}|$

\bigskip
\noindent The interpretation of the predicate $Gravity(Lorentz,Lorentz_{BPD},\mu_{gravity})$ is that for two $BPD$-conformally related spacetimes, for them to be {\em in-order} they have to be in $disorder$. Non-fuzzy $Gravity$ implies identical manifolds while fuzzy $Gravity$ admits different manifolds. So, ``$p$ is fuzzy iff $p$ is not fuzzy" reads dynamic perpetual oscillations (gravitational waves) of spacetime. These waves are continuous and perpetual and obviously much easier to phrase logically.
 They continue like this perpetually as the lessons of general relativity dictate a dynamic geometry of spacetime. Space and time lose order the more the speed approaches speed of light when they swap positions. $SySBPD$ is expressed as ``$p$ is fuzzy iff $p$ is fuzzy", where $p$ as in [1]: $p(t_1,t_2,\ldots,t_n,\mu)$. Or, ``$p$ is an atom of classical logic iff it is not an atom of classical logic". This is the $SpTBPD/SySBPD$ space/time vs. syntax/semantics mathematically representing the dynamic nature of spacetime as well as unifying it with quantum mechanics. $\kappa$ is the ``Bi-Polar Disorder Universal Constant" which can be observed by experiments relating {\em gravitational waves} to {\em quantum ones.}

\bigskip

\noindent Where the $FLP$ equations above predict a Bi-Polar Disorder dichotomy rather than the usual symmetry interpretation. $SpTBP$ naturally explains ``CP violation" as well as the 720 degrees for an electron to return to its state rather than 360 degrees (either {\em in-order} state or $disorder$, thus Bi-Polar Disorder). This explanation cannot be provided by the algebraic equation which only gives the mathematical prediction. The meaning of this formula that the two Lorentz manifolds
(perpetually) oscillate between two states. Once they are identical (classical logic programming),
the other with the deficit $\mu_{gravity}$ (FLP). It is impossible to
formulate this sort of oscillation as a wave against time; as usual
in physics. This new {\em logical} formulation resolves the problem.
$Lorentz$ and $Lorentz_{BPD}$ metrics describe two curved spacetimes
(spacetime and timespace) in $BPD$, thus (potentially) explaining
ripples in spacetime geometry. The {\em final paradox} is that $SySBPD$ is highly undesirable for computer science wouldn't be at all for physics.

\bigskip\bigskip
\noindent{\bf Discussion \& Conclusion:}

\bigskip

\noindent The following questions are addressed:

\begin{enumerate}

\item Can $SpTBPD$ be tested by experiment?

\item Does it make new predictions?

\item Does it generate new hypotheses?

\item Does $SpTBPD$ provide new explanations for strange observations?

\item $SySBPD$ implications to physics is $SpTBPD$ compared to late awakening to Godel's Incompleteness Theorem [1930-2002!!!]; in [2].

\item Is it a proposal for final theory?

\end{enumerate}

\bigskip
\noindent No theory is considered to be apodictically true unless supported by experimental results and observations. $SpTBPD$ is founded on general relativity and relativistic quantum mechanics. So it is obvious that the prediction of a ``Universe Bi-Polar Disorder Constant $\kappa$" relating spacetime/timespace from one side to $\Psi$ (particle)/$\Psi$ (anti-particle) from the other can be tested by experiment. In addition, several new hypotheses/new explanations have been (naturally) generated:

\begin{enumerate}

\item $SpTBPD$ regards gravitational waves (spacetime/timespace Bi-Polar Disorder) in curved space time as a result of fermion spin-like angular motion (in a Hilbert space) of flat spacetime and flat timespace.

\item The mathematical foundations of quantum mechanics (Hilbert spaces) could be unified with that of general relativity. The view of two Lorentzian manifolds in Bi-Polar Disorder ($BPD$-conformally related) can be restricted to two flat spacetime/timespace at quantum level. So, space/time dichotomy (spacetime vs. timespace) at both the super-galactic and the sub-atomic levels.

\item A quantum interpretation of the Big Bang and the expansion of the universe due to the new dichotomy of particle/anti-particle vs. spacetime/timespace.

\item Extending the Pauli exclusion principle from the sub-atomic level to the super-galactic.

\item When the (poor) author first learnt of the 720 degrees for an electron to return to its original state, it was no surprise unlike many others. This is a natural $SpTBPD$ quantum view of {\em in-order} and $disorder$ states.

\item When the (poor) author learnt that ``CP violation" is not complete symmetry, he felt that was absolutely normal and consistent with $SpTBPD$. $SpTBPD$ predicts Bi-Polar Disorder in nature rather than symmetry. But in order to maintain the order, there has to be disorder (from the Big Bang to the expansion of the Universe), resulting whenever the symmetry attempts to become complete, it could recur somewhere else incomplete.

\end{enumerate}

\noindent The answers are positive for $SpTBPD$: It makes a new prediction $\kappa$ that can be found by experiment and it provides new hypotheses. Not only this, but also it provides (natural) and consistent explanations for unexplained phenomenon. The story about the implications of the celebrated G$\ddot{o}$del's Incompleteness theorem is indeed a sad one as detailed in the paper by Reverend Father Professor Stanley L. Jaki [2]. A more bizarre story is expected for {\em fuzzy logic programming} where new dichotomies have been identified and mathematically related to well-established ones: Syntax/Semantics vs. Spacetime/Timespace vs. Particle/Anti-particle vs. {\bf Wave/Particle Bi-Polar Disorder}, not Wave/Particle duality!

\bigskip
\noindent Whether it is a proposal for a final theory, the answer is simply ``No". In [2], it has been confirmed that even after considering G$\ddot{o}$del's Incompleteness theorem's implications to physics, a final theory is possible; but it is not possible to prove this fact rigourously. But G$\ddot{o}$del's Incompleteness theorem was for Peano's arithmetic, i.e. natural numbers. Physics must employ real numbers. So, if Peano's arithmetic has infinite number of axioms, the author presents the hypothesis that a final theory wouldn't need only infinite number of axioms, but also a mathematical language whose alphabet is infinite!

\bigskip
\noindent{\bf References:}

\begin{enumerate}

\item Rafee Ebrahim Kamouna, ``Fuzzy Logic Programming", Fuzzy Sets and Systems, 1998.

\item Stanley L. Jaki:``A Late Awakening to G$\ddot{o}$del in Physics", pirate.shu.edu/

~jakistan/JakiGodel.pdf, accessed 15/02/2008.

\item Stephen W. Hawking, ``G$\ddot{o}$del and the End of Physics", www.damtp.cam.ac.uk/

strtst/dirac/hawking, accessed 15/02/2008.

\item http://en.wikipedia.org/wiki/Causal\_structure.

\item http://en.wikipedia.org/wiki; keywords: ``Introduction to General Relativity" and ``General Relativity".

\end{enumerate}

\end{document}